\RequirePackage{fix-cm}
\documentclass[twocolumn]{svjour3}          
\smartqed  
\usepackage{graphicx}
\usepackage{amsmath}
\usepackage{amsfonts}
\usepackage{amssymb}
\usepackage[titletoc,title]{appendix}
\usepackage{color}

\newcommand{\fref}[1]{Fig.~\ref{#1}}
\newcommand{\sref}[1]{Sec.~\ref{#1}}
\newcommand{\aref}[1]{Appendix~\ref{#1}}
\newcommand{\eref}[1]{Eq.~(\ref{#1})}

\begin{document}
\title{Quantum Noise from a Bohmian perspective: fundamental understanding and practical computation in electron devices}

\author{D. Marian, E. Colom\'{e}s, Z. Zhan  and  X. Oriols }
\institute{E. Colom\'{e}s, Z. Zhan and  X. Oriols \at
              Departament d\rq{}Enginyeria Electr\`{o}nica, Universitat Aut\`{o}noma de Barcelona, Spain. \\
              \email{xavier.oriols@uab.es}          
           \and
           D. Marian \at
              Dipartimento di Fisica dell'Universit\`a di Genova and INFN sezione di Genova, Italy.}

\date{\today}

\maketitle
\begin{abstract}
In the literature, the study of electron transport in quantum devices is mainly devoted to DC properties. The fluctuations of the electrical current around these DC values, the so-called quantum noise, are much less analyzed. The computation of quantum noise is intrinsically linked (by temporal correlations) to our ability to understand/compute the time-evolution of a quantum system that is measured several times.  There are several quantum theories that provide different (but empirically equivalent) ways of understanding/computing the perturbation of the wave function when it is  measured. In this work, quantum noise associated to an electron impinging upon a semitransparent barrier is explained using Bohmian mechanics (which deals with wave functions and point-like particles). From this result, the fundamental understanding and practical computation of quantum noise with Bohmian trajectories are discussed. Numerical simulations of low and high frequency features of quantum shot noise in a resonant tunneling diode are presented (through the BITLLES simulator), showing the usefulness of the Bohmian approach.  
\end{abstract}

\section{Introduction}
\label{sec1}

Historically, the definition of noise was related to the sound: A noise is an unwanted, unpleasant and confusing type of sound.\footnote{In fact, the word noise is etymologically derived from the Latin word \emph{nausea}, meaning seasickness.} However, such definition is ambiguous. What does it mean unwanted, unpleasant or confusing? An attempt to provide a more academic definition comes from music: Noise is a non-harmonious or discordant group of sounds. Again, however, the definition is not free from ambiguities because one man's noise is another man's music \cite{Landauer1}. 

A more scientific definition closer to our interest in electrical devices comes from communications: A noise is an electric disturbance that interferes with or prevents reception from a signal or information. For example, the buzz in a telephone call. Thus, we realize that once we have a precise definition of what is a signal, the meaning of what is noise becomes perfectly clear: It is the difference between the measured value and the signal.  

\subsection{Quantum noise in electrical devices from an experimental point of view}
\label{sec1.1}

As discussed above, the answer to what is noise in electrical devices depends on our definitions of the electrical signal. For most DC applications, the signal is just a time average value of the current. For frequency applications, the signal is equivalently defined as a time average value, but using a shorter time interval (related to the inverse of the operating frequency). In other applications, mainly digital applications, the signal is related to a time average value of the voltage in a capacitor. Hereafter, we will assume that the electrical signal is the DC value of the current, referenced by the symbol $\langle I \rangle$.  All fundamental and practical issues discussed here for the DC signal (and its noise) can be easily and straightforwardly extended to those other types of electrical signals. 

What is measured in a laboratory for the DC signal is the time average value of the instantaneous current $I(t)$ in a unique device during a large period of time $T$:
\begin{equation}
\langle I \rangle = \lim_{T\rightarrow \infty}\frac{1}{T}\int^{T}_0 I(t) dt.
\label{noise1}
\end{equation}
Once we have defined the signal $\langle I \rangle$ as the DC value, in principle, the noise can be quantified by time averaging the difference between the measured value of the current $I(t)$ and the signal in a unique device:
\begin{equation}
\triangle I^2 = \lim_{T\rightarrow \infty}\frac{1}{T}\int^{T}_0 (I(t)-\langle I \rangle)^2dt.
\label{noise2}
\end{equation}
The square of the difference avoids positive and negative cancellations.\\

At this point, it is very important to realize that $I(t)$ presents very rapid fluctuations that cannot be captured by the standard laboratory apparatuses. Any experimental setup that measures the current fluctuations  behaves as a low-pass filter (i.e. the current fluctuations at frequencies higher than the apparatus cut-off frequency are not measured). Therefore, the experimentally accessible information about the current fluctuations is not given by \eref{noise2}, but by the power spectral density of the fluctuations $S(w)$ (and its related magnitudes). From the Wiener-Khinchine relation, the  power spectral density can be defined as the Fourier transform of the time average definition of this autocorrelation function $\triangle R(\tau)$ :
\begin{equation}
\triangle R(\tau) =\lim_{T\rightarrow \infty}\frac{1}{T}\int^{T}_0 \triangle I(t_1) \triangle I(t_1+\tau) dt_1,
\label{noise3}
\end{equation}
 where $\triangle I(t)=I(t)-\langle I \rangle$. A straightforward calculation shows that \eref{noise3} can be rewritten as $\triangle R(\tau) = R(\tau)-\langle I \rangle^2$ with:
\begin{eqnarray}
R(\tau) = \lim_{T\rightarrow \infty}\frac{1}{T}\int^{T}_0 I(t_1) I(t_1+\tau) dt_1.
\label{noise4}
\end{eqnarray}
Then, the Fourier transform of \eref{noise3} gives the noise power spectral density $S(w)$:
\begin{eqnarray}
S(w) = \int_{-\infty}^{\infty} \triangle R(\tau) e^{-jw\tau} d\tau.
\label{noise5}
\end{eqnarray}
It is quite trivial to realize that the definition of the spectral density $S(w)$ in \eref{noise5} and \eref{noise3} is consistent with the definition of the total noise\footnote{Technically, $S(w)$ defined in \eref{noise5} is non-negative and symmetric with respect to $w$. Then, since only positive frequencies $w$ are measured in a laboratory, the measured density includes our $S(w)$ and $S(-w)$, and the integral of the noise spectrum measured in a laboratory runs from $0$ till $\infty$.} in \eref{noise2}:
\begin{eqnarray}
\triangle I^2 = \int_{-\infty}^{\infty} S(w) dw,
\label{noise6}
\end{eqnarray}
where we have used the definition of the delta function $\delta(\tau)=\int_{-\infty}^{\infty} e^{-jw\tau} dw$.  \\

It is very relevant for the rest of the paper to realize that the measurement of $S(w)$ through the function $R(\tau)$ defined in \eref{noise4} requires the knowledge of the measured value of the current during all $t$. Thus, we have to make predictions about the evolution of the electronic device while being (continuously) measured. In a classical scenario, such discussion about measurement is generally ignored. On the contrary, for quantum systems, it has very relevant implications because the evolution of a system with or without measurement can be dramatically different. 

If the electronic device satisfies the ergodic theorem \cite{Price,ergodic}, a continuous measurement of the system can be avoided in practical computations. Let us see in what sense ergodicity can simplify our noise computations. In general, the \emph{mathematical} concept of a random process is used to deal with noise.  A random process requires a sample space. In our case, we can define an ensemble of \emph{identical} electrical devices\footnote{At this point, the reader will wonder that, in typical laboratory experiments, only one electronic device is available (not an ensemble of them). Then, as a practical definition of ensemble, we can define the instantaneous current measured in different time-intervals: $I^{\gamma_1}(t)$ for the instantaneous current measured during the first time interval, $I^{\gamma_2}(t)$ for the second interval, and so on. }, each one labeled by the sample space variable $\gamma$. Then, the (instantaneous) current is labeled by the random process $I^{\gamma}(t)$. For a fixed time, $t_1$, the quantity  $I^{\gamma}(t_1)$ is a random variable. For a fixed device $\gamma_1$, the function $I^{\gamma_1}(t)$ is a well-defined non-random function of time. Finally,  $I^{\gamma_1}(t_1)$ is just a real number. Often the sample space variable $\gamma$ is omitted in the notation. The DC value of the current in \eref{noise1} can be alternatively defined for an ergodic system as:
\begin{equation}
\langle I \rangle = \sum_{i} I_{i}(t_1) P(I_i(t_1)),
\label{noise7}
\end{equation}
where $P(I_i(t_1))$ is the probability of getting $I_i$ at time $t_1$. These probabilities are defined as the ratio of the number of devices providing $I_i$ divided by the total number of devices. It is important to realize that the experimental evaluation of \eref{noise7} requires only one measurement of the current at $t_1$ in a large number of \emph{identical} $\gamma$-devices. Then, the theoretical predictions of \eref{noise7} do only need to determine the free (without measuring apparatus) evolution of the electronic device from the initial time $t_0$ till $t_1$. See a detailed discussion in appendix \ref{app-noise} about how ergodicity avoids the complications of the measurement in a quantum system. Obviously, we can compute the total noise represented in \eref{noise2} from a unique measurement in ergodic systems: 
\begin{equation}
\triangle I^2 = \sum_{i} (I_{i}(t_1)-\langle I \rangle )^2 P(I_i(t_1)),
\label{noise7bis}
\end{equation}
However,  the noise measured in a laboratory is not given by $\triangle I^2$, but by $S(w)$ in \eref{noise5}. We repeat the reason explained in \eref{noise2}. The amount of noise generated by an instantaneous current evolving for example from $I(t_1)=5$ mA to $I(t_2)=10$ mA during a time interval of $t_2-t_1=\tau=1$ fs,  is not captured from the state-of-the-art laboratory apparatuses (which already have difficulties to capture noise at frequencies higher than a few of Terahertzs).   From an experimental point of view, in fact, it is easy to get $S(w\rightarrow 0)$, but impossible to get $S(w\rightarrow \infty)$. We can compute the noise power spectral density $S(w)$ from the ensemble average version of the autocorrelation defined in \eref{noise4} as:
\begin{eqnarray}
R(t_1,t_2)=\sum_i\sum_jI_j(t_2)I_i(t_1)P\big(I_j(t_2),I_i(t_1)\big).
\label{noise8}
\end{eqnarray}
In general, we can assume that the instantaneous current in an electronic device behaves as a wide-sense stationary random process. Then,  $\langle I \rangle$ in \eref{noise7} is constant and time-independent. Identically, then, the autocorrelation function in \eref{noise8} depends only on the time difference $ R(t_1,t_1+\tau)= R(\tau)$ with $t_2=t_1+\tau$. Finally, we use \eref{noise5} with $\triangle R(\tau) = R(\tau)-\langle I \rangle^2$ computed from \eref{noise8}, to get the noise power spectral density $S(w)$. 

It is important to emphasize (for a posterior discussion) that the probability $P\big(I_j(t_2),I_i(t_1)\big)$ implies a two-measurement process for each electronic device. The system evolves freely (without interaction with the measurement apparatus) from $t_0$ till $t_1$ when the current is measured,  giving the value $I_i$. Then,  the system evolves freely again until time $t_2$, when the system is measured again giving $I_j$.  In summary, even if the ergodicity argument is invoked, the noise computation through the autocorrelation function requires, at least,  two measurements at different times in a single device (and the average over all $\gamma$-devices). We anticipate that our computations with Bohmian mechanics will not assume ergodicity (which is not an obvious property for open systems out of equilibrium \cite{Price}), but the prior expressions requiring a continuous measurement of the current.  

Let us emphasize that the previous discussion is valid for either classical or quantum devices. The adjective \emph{quantum} emphasizes that the signal and the noise are computed or measured in an electrical device governed by quantum laws \cite{Landauer2,Beenakker1,Beenakker2,Buttiker1}.  If the electronic device is not ergodic,  expression (\ref{noise4}) requires a continuous measurement of the current $I(t)$. On the contrary, for an ergodic electron device, expression (\ref{noise7}) requires one unique measurement, while expression (\ref{noise8}) requires a two-times measurement when dealing with the power spectral density $S(w)$. 

Up to here, we realize that the definition of quantum noise seems very trivial. Then, why does the concept of quantum noise have a  halo of mystery around it?

\subsection{Quantum noise in electrical devices from a computational point of view}
\label{sec1.2}

Our previous definition about what is quantum noise does not answer the question of how we compute it. If we want to predict the values $I(t)$ used in \eref{noise2}  and \eref{noise4} or the probabilities $P(I)$ and $P\big(I_j(t_2),I_i(t_1)\big)$ for \eref{noise7} and \eref{noise8}, we require a quantum theory. 

There are several quantum theories available in the literature that, by construction, are empirically equivalent when explaining all quantum phenomena. Among others, the so-called Copenhagen or orthodox interpretation \cite{Copenhagen1,Copenhagen2}, Bohmian mechanics \cite{Bohm,OriolsBook,NinoBook}  or the many-worlds theory \cite{Everett}. Any theory has usually two different planes. First, the formalism, which is a set of mathematical rules (using elements such as wave functions, operators, trajectories) that allow
 us to make practical computations that reproduce experimental results. The formalism of a theory provides an answer to the question: How do we compute quantum noise? The second plane of a theory is its interpretation. It tries to provide a deep connection on how the mathematical rules and its elements explain how nature works. The interpretation of the theory provides answers to the question: Which is the physical origin of quantum noise? Each quantum theory will provide its own answers to both questions. 

Many people argue that the only important part of a quantum theory (once we know it is empirically valid) is its formalism because it is the only part we need to make computations. Certainly, one can make noise computations using any of the available formalisms without worrying about its interpretation. At the end of the day, by construction, each theory should give the same predictions. Other people argue that even when one is only interested in computations, a correct understanding of the interpretational issues of each theory is fruitful because it provides an enlarged vision about how correctly apply the theory in unsolved problems (abandoning the \emph{shut up and calculate} \cite{SUAC}). We will return to this very point later, at the conclusions in \sref{sec5}.  

Now, we want to clarify why quantum noise is specially sensible to fundamental quantum mechanical issues. Any electrical device (or any experiment) is connected to a measuring apparatus. In our case, an ammeter to get the electrical current. Quantum noise is sensible to the (ammeter) measuring process. As stated in \eref{noise8}, in order to obtain the noise, the quantum system has to be measured, at least, twice. This two-time measurement faces directly with one of the most complex issues in quantum mechanics, namely, which is the perturbation of the quantum wave function when a measurement is performed. Historically, this perturbation is known with the somehow scary name of \emph{the collapse of the wave function}. Can we ignore it? Definitively not if temporal correlations need to be correctly predicted. See for example, Ref. \cite{Buttiker3}:\emph{\lq\lq{}The fluctuations ... are a consequence of a probabilistic reflection and transmission probability (a wave phenomena) and are a consequence of the fact that detectors register either a transmitted or a reflected particle (a particle phenomena)\rq\rq{}}. The measurement process is hidden in the word \emph{detectors}. 

We also mention that the fundamental understanding/computing of the measurement process can be largely relaxed when dealing with DC predictions. They can be computed from an ensemble of devices with only one measurement in each device, so we can ignore the evolution of the quantum system after the measurement. See \aref{app-noise} to enlarge this point.

In this paper, we will provide an explanation to the role of collapse in quantum noise from a Bohmian perspective. We emphasize that we are not saying that the Bohmian answer is the best one. Answers from other theories are equally satisfactory, and provide the same predictions. We are just defending that it is a consistent answer that in the authors\rq{} opinion provides a quite intuitive and understandable explanation of quantum noise and also a numerically accessible formalism. In \sref{sec2} we explain how the Copenhagen interpretation explains the multi-time measurement process in a experiment with a flux of electrons impinging upon a tunneling barrier, by introducing the notion of operators. In \sref{sec3} we provide an explanation of the same experiment using Bohmian mechanics, without using operators. Then, in \sref{sec4} we illustrate how the formalism of Bohmian mechanics exposed in \sref{sec3} can be applied in practical problems to calculate the quantum noise in electrical devices, including Coulomb and  exchange interactions. Finally, we conclude in \sref{sec5} explaining how the different theories explain the origin of quantum noise.

\section{Multi-time measurement with operators}
\label{sec2}

A typical scenario when discussing quantum noise in electrical devices is a flux of electrons impinging upon a partially transparent barrier (located in the middle of the active region).  Electron transport through the barrier takes place by tunneling. Electron is either transmitted or reflected, but not both! \cite{Beenakker1,Buttiker2,Landauer3} We get a transmitted electron with a probability $T$, while a reflected one with probability $R=1-T$.  To simplify the discussion, we consider a constant injection of electrons (at zero temperature), one by one.  Each electron, after measurement at time $t_1$, will appear randomly at the left or at the right of the barrier. The time averaged number of transmitted electrons will be proportional to $T$, but the number of transmitted electrons fluctuates instantaneously because of the randomness of the transmission. These fluctuations of the number of transmitted electrons (when compared with the DC signal) are named partition noise \cite{Beenakker2,Buttiker3,Landauer3}.

There are many other sources of noise in electrical devices, for example, the $1/f$ noise which become very relevant at low frequencies \cite{Beenakker2,Buttiker1}. In this paper, we will only deal with partition noise due to a tunneling barrier. In \sref{sec4}, we will discuss partition noise considering also the injection of electrons at a finite temperature (the so-called thermal noise). The fluctuations due to both processes simultaneously are known in the literature as shot noise  \cite{Landauer1,Beenakker1,Buttiker1,Levitov}.

\begin{figure}
      \includegraphics[width=0.97\columnwidth]{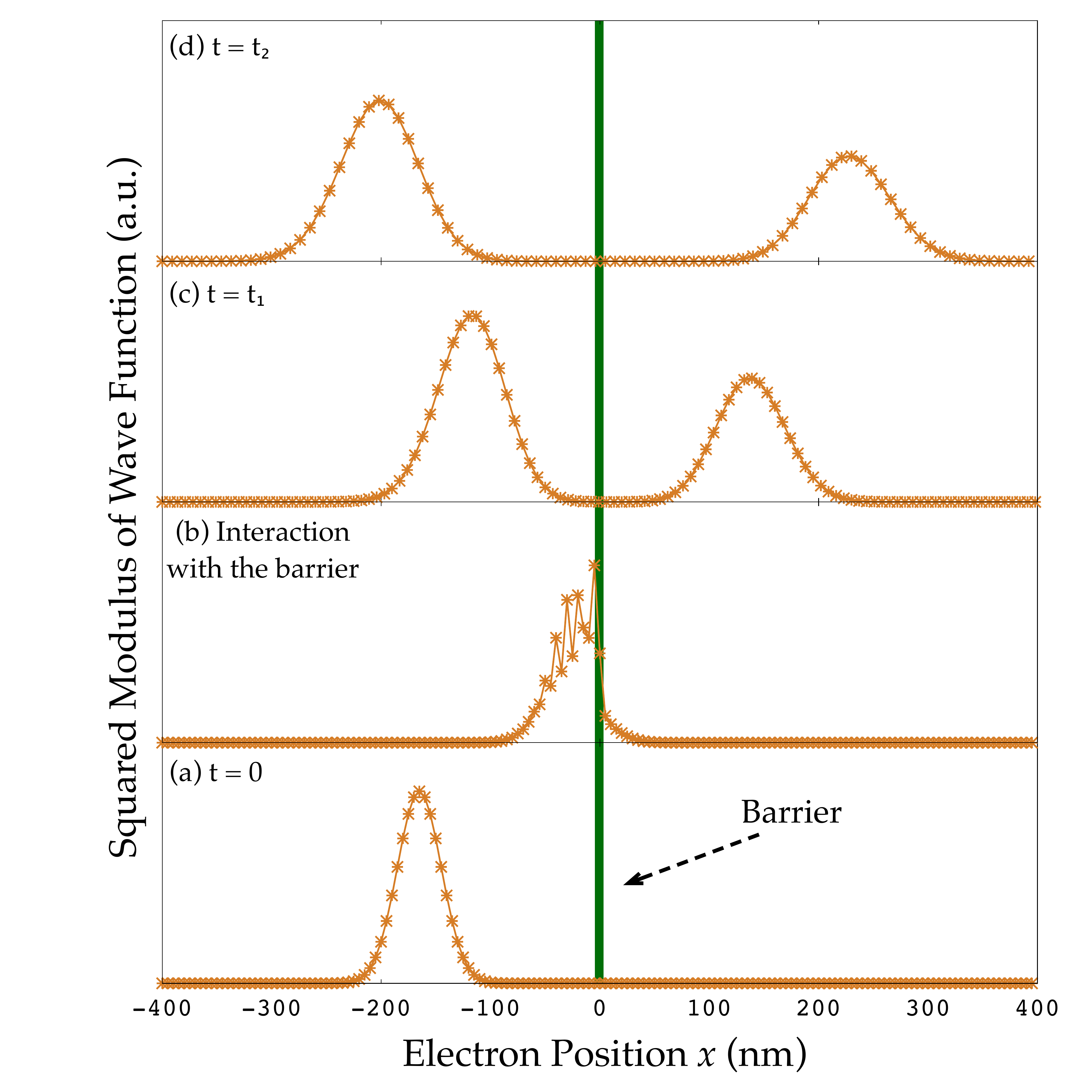}
       \caption{Evolution of the squared modulus of the wave function of an electron impinging on a tunneling barrier (green solid line). We plot four different times corresponding to (a) initial time, (b) the moment when the wave function interacts with the barrier, (c) the time $t_1$ when it occurs the first measurement and (d) time $t_2$ corresponding to the second measurement. At time $t_1$ and $t_2$, because of the unitary evolution, the electron can be detected at both sides of the barrier (Color figure online)}
        \label{figure1}
\end{figure}

\begin{figure}
      \includegraphics[width=0.97\columnwidth]{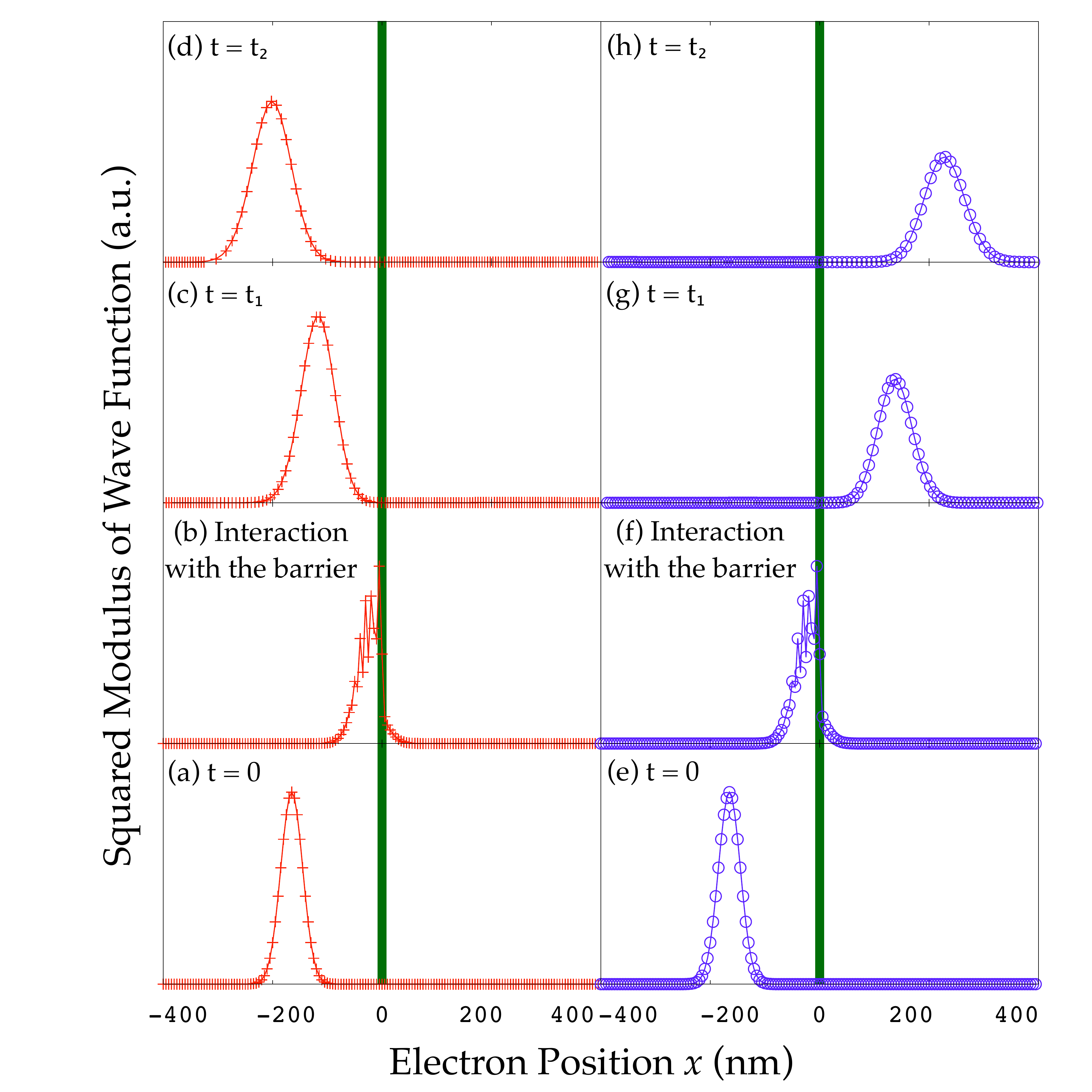}
       \caption{(a), (b), (c) and (d) Non-unitary evolution of the wave function for a reflected electron detected at time $t_1$ at the left side. (e), (f), (g) and (h) Non-unitary evolution of the wave function for a transmitted electron detected at time $t_1$ at the right side. Symbols are the same as in \fref{figure1} (Color figure online)}
        \label{figure2}
\end{figure}

In this section we discuss how the partition noise is understood within the orthodox interpretation of quantum mechanics, also known as Copenhagen interpretation \cite{Copenhagen1,Copenhagen2}. Let us specify that most available formulations of shot noise are developed within this orthodox interpretation \cite{Landauer1,Beenakker1,Beenakker2,Buttiker3,Buttiker2,Landauer3}. We consider a very simple example, but with a detailed discussion of the role played by the measuring apparatus (the ammeter). The Copenhagen interpretation associates a wave function $\Psi(\bar{x}_N,t)$ to a system of $N$ particles. In principle, such wave function \emph{lives} in a $3N + 1$ dimensional configuration space. Within the first non-relativistic quantization language, the evolution of this wave function is defined by two laws \cite{Cohen}. The first law, known as Schr\"odinger equation, states that (when the system is not measured) the wave function evolves unitarily and deterministically according to the following equation

\begin{equation}
i\hbar \frac{\partial \Psi(\bar{x}_N,t)}{\partial t}=H\Psi(\bar{x}_N,t),
\label{TDSE}
\end{equation}

where $H=\big[\sum_i-\frac{\hbar^2}{2m_i}\nabla^2_i+U(\bar{x}_N,t)\big]$. With $U(\bar{x}_N,t)$ we denote a generic interaction potential in the position representation, with $m_i$ the mass of the $i$-th particle and with $\bar{x}_N = (x_1,x_2,..., x_N)$ the multidimensional vector in the configuration space. To provide a simple discussion of the partition noise in a tunneling barrier, let us assume that each electron in our experiment can be described by a single-particle wave function (we neglect the exchange and the Coulomb interaction among electrons). In Fig.~\ref{figure1} we plot the (unitary) evolution of such wave function solution of \eref{TDSE}. Is the (unitary) Schr\"odinger equation alone depicted in Fig.~\ref{figure1} enough to understand quantum noise? No. The orthodox theory has a second law, known as the \emph{collapse} of the wave function, that takes into account the effects of the interaction of a measuring apparatus with the quantum (sub)system. This second law can be found in many textbooks \cite{Cohen}.  It requires a new non-unitary operator $A$. This operator is different from the unitary Schr\"odinger evolution, which is generated by the Hamiltonian seen below \eref{TDSE}, and it must be able to encapsulate all the interactions of the quantum systems with the rest of the particles (including the ammeter, the cables, the environment, etc). This new operator $A$ is the only tool provided by the theory to determine the possible results of a measurement. In principle we do not know anything about this operator except that it is a (hermitian $A = A^{\dagger}$) function whose (real) eigenvalues $a_n$ of its spectral decomposition are the possible results of the measurement. Once the system in Fig.~\ref{figure2} is measured (and not before), the wave function is projected to one of the eigenstates of the mentioned operator in a non-unitary evolution.\footnote{The measurement described in most textbooks is called ``projective" (\emph{strong}) measurement. There exists, for example, another type of measurement known as \emph{weak} measurement, which is useful to describe situations where the effects of the apparatus on the measured system is just a small perturbation.} After the collapse, the \textit{new} wave packet evolves again according to the time-dependent Schr\"odinger equation until a new measurement is done. 

For simplicity, in our present conceptual discussion let us assume a reasonable (but ad-hoc) operator (why this operator is reasonable will be clarified in \sref{sec3}). Such operator provides the following perturbation of the wave function. If the electron is \emph{randomly} measured as a reflected electron at $t_1$, the transmitted part of the wave function is eliminated. This measuring process corresponds to Figs. \ref{figure2} (c) and (d) where only the reflected wave function survives after $t_1$. Equivalently, the measurement process associated to \emph{randomly} getting a transmitted electron corresponds to eliminating the reflected part, as seen in Figs. \ref{figure2} (g) and (h). 

Now, by comparing the evolutions of the wave functions in \fref{figure1} and \fref{figure2}, it is obvious that the former is wrong. By looking at \fref{figure1}, it could be the case that an electron found at time $t_1$ at the right (transmitted) can be found in a later time $t_2$ at the left as a reflected electron (see the evolution of the probability density in \fref{figure1}). This sequence of possibilities is wrong. Experimental results confirm that  once, say time $t_1$, the electron is detected at one side, in a later time $t_2$ it is always found at the same side.  Then, we get a very valuable lesson from the Copenhagen explanation: the (unitary) Schr\"odinger equation alone is not able to explain completely quantum noise. It is necessary to include the collapse of the wave function to understand properly what is quantum noise (temporal correlations). The popular arguments that \emph{\lq\lq{}Shot noise is a consequence of quantization of charge\rq\rq{}} \cite{Buttiker1} or \emph{``This is the noise that arises from the graininess of the current''} \cite{Landauer3} emphasize exactly this very point.  

All mentioned orthodox formalisms dealing with quantum noise reproduce experimental results successfully because they include the measurement process inside \cite{Landauer1,Beenakker1,Beenakker2,Buttiker3,Buttiker2,Landauer3,Levitov}.   
Most of them do not discuss explicitly which is the operator associated with the ammeter.  
Over the years, physicists have identified the operators, by developing instincts on which are the effects of measurements in the wave function. There are scenarios (as the one depicted in \fref{figure2}) where it is quite obvious which operator is the \emph{right} one. On the contrary, for example, when measuring the total (conduction plus displacement) current it is not at all obvious which are the relevant operators. Is this measurement process \emph{continuous} or \emph{instantaneous}? Does it provide a \emph{strong} or a \emph{weak} perturbation of the wave function? The answers to these questions are certainly not simple. The Copenhagen theory itself does not answer these \emph{technical} questions about how to find the \emph{right} operator. Can other quantum theories provide additional help?

\section{Multi-time measurement without operators}
\label{sec3}

In the previous section, we discussed how the Copenhagen formalisms can be successfully used to understand quantum noise in electrical devices. One technical difficulty with this formalism is the proper definition of the \emph{right} operator that determines the collapse of the wave function when, for example, the total (conduction plus displacement) current is measured. 

There are alternative theories which account for the perturbation of the wave function during a measurement process in a different way, without operators. The one that we will develop here is Bohmian mechanics. Let us emphasize again that both (Copenhagen and Bohmian) theories are empirically equivalent so that the preferences of one in front of the other are related to computational abilities, intuitive results, etc. \cite{OriolsBook}.

In the Bohmian theory, the complete description of a quantum system of $N$ particles is given by the (same) wave function, $\Psi(\bar{x}_N,t)$ mentioned in \sref{sec2}, and by the actual positions of the point-like particles, $\bar{X}_N(t) = (X_1(t), X_2(t), ..., X_N(t)) $\footnote{We denote with capital letter $X$ the actual position of the particle, while the lower case letter $x$ is used to indicate generic positions. With the barred letter we refer to a multidimensional vector in the configuration space, while a letter without bar denotes the 3-dimensional vector in physical space.} (see \aref{app-bohm} and Refs. \cite{OriolsBook,NinoBook}  for a more detailed discussion on this theory).
We emphasize that the evolution law for the wave function $\Psi(\bar{x}_N,t)$ is the same as in standard quantum mechanics: the Schr\"odinger equation (\eref{TDSE}). The wave function \emph{guides} the movement of the actual positions of the particles in time, according to the so-called \emph{guidance equation}, which defines the velocity of the $i$-th particle as

\begin{eqnarray}
v_i(t) &=& \frac{J_i(\bar{x}_N,t)}{|\Psi(\bar{x}_N,t)|^2}\Big|_{\bar{x}_N=\bar{X}_N(t)} = \nonumber \\
&=& \frac{\hbar}{m_i} \text{Im} \frac{\nabla_i \Psi(\bar{x}_N,t)}{\Psi(\bar{x}_N,t)}\Big|_{\bar{x}_N=\bar{X}_N(t)},
\label{eq-guidance}
\end{eqnarray}

where $J_i(\bar{x}_N,t)$ is the usual probability current density, defined as $J_i(\bar{x}_N,t) = \frac{i\hbar}{2m_i}\left(\Psi \nabla_i \Psi^*-\Psi^* \nabla_i \Psi \right)$, associated to the $i$-particle, $|\Psi(\bar{x}_N,t)|^2$ is the usual probability distribution and Im denote the imaginary part. We note that \eref{eq-guidance} describes the \emph{velocity field} for the $i$-particle and depends on the actual position of all particles of the system $\bar{X}_N(t)$. Each particle follows a definite trajectory which can be obtained integrating in time the velocity field     

\begin{equation}
X_i(t) = X_i(0) +\int_0^t v_i(t') dt',
\label{eq-position}
\end{equation}

where $X_i(0)$ is the initial position of particle $i$.\\
A proper ensemble of these trajectories (proper means that the initial position of each trajectory of the ensemble is selected according to the initial squared modulus of the wave function, see \eref{QEH} in \aref{app-bohm-3}) reproduces the time-evolution of the many-particle wave function at any later time. \\
In \sref{sec2}, we saw that in order to reproduce the experimental results, we have used the notion of operators to describe how the wave function of a measured system is modified under a measurement process. In the Bohmian theory, we simply consider the apparatus as another (big and complex) quantum system interacting with our measured system. The interaction among them is then included in the Hamiltonian of \eref{TDSE} as any other interaction. Then from the unitary evolution of the many-particle wave function (system plus apparatus) we can look at the behavior of the wave function of the measured system. The latter is called \emph{conditional wave function} (an exclusive concept belonging to Bohmian mechanics) and it is defined from the many-particle wave function in the configuration space, fixing all the actual particles positions excluding that of our subsystem (see \aref{app-bohm-2} for more details).

\begin{figure}
\centering
\includegraphics[width=0.97\columnwidth]{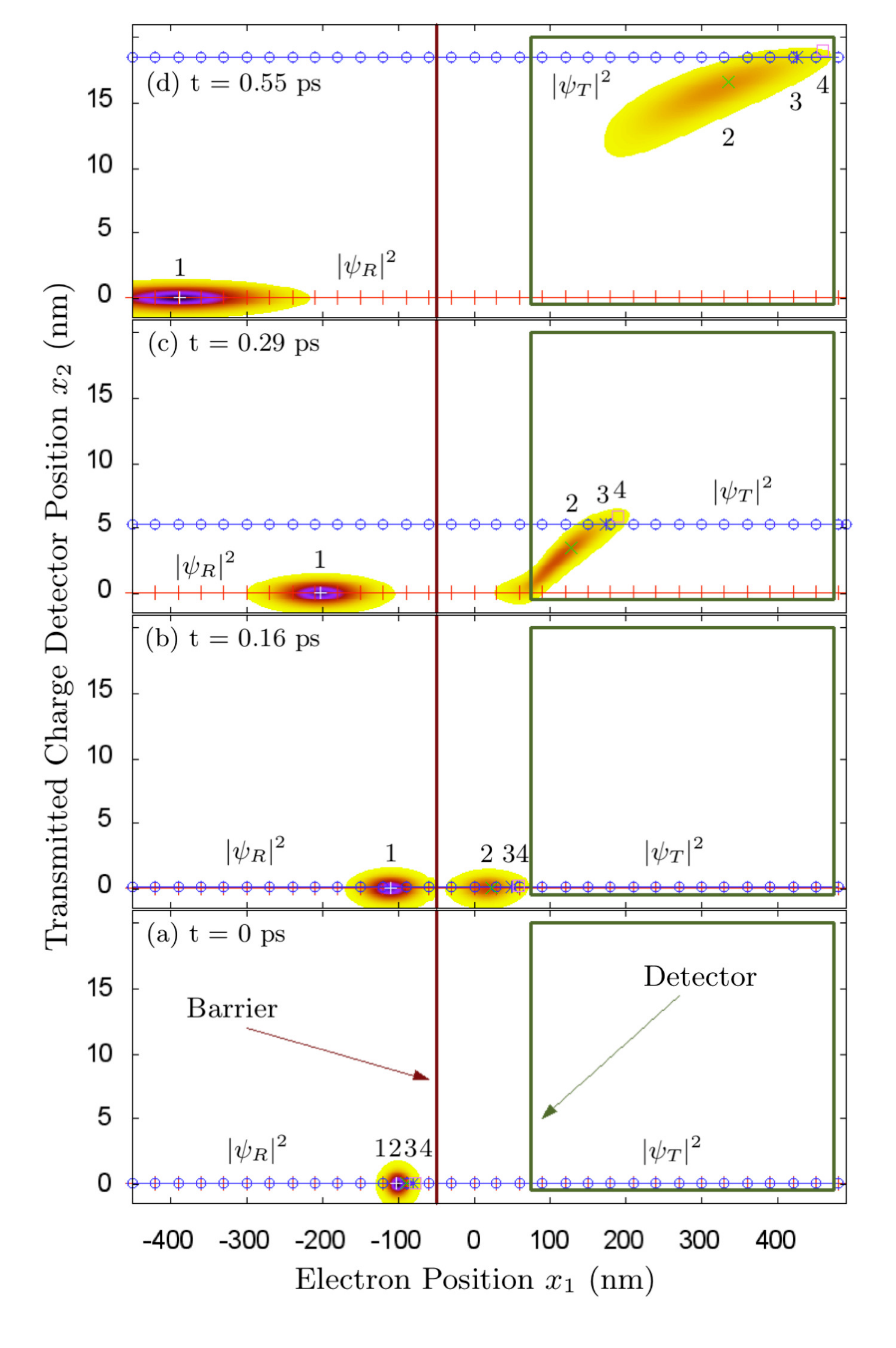}
\caption{Time evolution of the squared modulus of $\Psi(x_1,x_2,t)$ at four different times. The configuration space region where the \emph{transmitted charge detector} is present is indicated by a rectangle and the barrier by a solid line. The $+$ line indicates the modulus of the conditional wave function $|\psi_R|^2=|\Psi(x_1,X_2^{\alpha=1}(t),t)|^2$, while the  $\odot$ line corresponds to $|\psi_T|^2=|\Psi(x_1,X_2^{\alpha=3}(t),t)|^2$. Four trajectories $\{X_1^\alpha(t),X_2^\alpha(t)\}$ with different initial positions are presented with $\square$, $*$, $\times$ and $+$. The actual detector position associated with the reflected trajectory ($+$) with $\alpha=1$ does not move because there is no interaction between this trajectory and the detector (Color figure online)}
\label{conditional wave function-figure1}
\end{figure}

Let us provide a quite realistic (in particular, non-instantaneous, but in some ways schematic) example in which we can numerically track the behavior of the conditional wave function during the measurement process of the partition noise discussed in \sref{sec2}. The quantum system is an electron labelled as $X_1$ impinging on an
external tunneling barrier. Behind the barrier there is a measuring device, that we call ``transmitted charge detector" modeled as a single degree of freedom $X_2$ (thought as the center of mass of a complex system), which can detect the successful transmission of an electron. First, we have an interaction of the electron with the potential barrier and, subsequently, an interaction with the transmitted charge detector. It is important to stress that both interactions are regarded at the very same level within Bohmian mechanics. The measurement interaction introduces a channelling of the wave function in the configuration space such that the desired property of the ``quantum system'' (here, whether the electron is reflected or transmitted) can be read off from the final position $X_2$ of a particle, thought of as the pointer of the apparatus. 
The interaction between the electron and the pointer can be modeled as:
\begin{eqnarray}
H_{int}= \lambda Q(x_1) P_{x_2} = -i \hbar\lambda Q(x_1) \frac {\partial } {\partial x_2},
\label{eq-interaction2} 
\end{eqnarray}
where $P_{x_2} = -i\hbar \partial /\partial x_2 $ is the momentum operator of the detector and $\lambda =50\; nm/ps $ is the interaction constant. $Q(x_1)$ is a function that is equal to zero when the electron is outside the detector, ($x_1 < 75 \; nm$ in \fref{conditional wave function-figure1}), and is equal to one when the particle is inside the detector ($x_1 > 75 \; nm$).\footnote{The transition of $Q(x_1)$, from zero to one, is done softly in order to minimize the perturbation of the ``quantum system'' as explained in \cite{albareda}.}  In \fref{conditional wave function-figure1} the region in the configuration space in which this function is different from zero is represented by a rectangle.
The many-particle Schr\"odinger equation reads
\begin{eqnarray}
 &&i \hbar \frac{\partial \Psi(x_1,x_2,t)}{\partial t}=
 \Big(- \frac{\hbar^2}{2m}\frac {\partial^2} {\partial x_1^2}  - \frac{\hbar^2}{2 M}\frac {\partial^2} {\partial x_2^2}  +  \nonumber \\
 &+& U(x_1) - i \hbar \lambda Q(x_1) \frac {\partial} {\partial x_2}\Big) \Psi(x_1,x_2,t), \; \; \;
\label{eq-schr2D}
\end{eqnarray}
where $m$ is the effective mass of the electron, $M$ is the mass of
the apparatus pointer and $U(x_1)$ is the external potential energy barrier.

The main feature of a transmitted charge detector is that the center of mass of the wave function in the $x_2$ direction has to move if the
electron is transmitted and it has to be at rest if the electron is reflected.  We solve \eref{eq-schr2D} numerically considering as initial wave function the products of two gaussian wave packets, i.e. $\Psi(x_1,x_2,0) = \psi(x_1,0)\phi(x_2,0)$. All details of this simulation can be found in \cite{albareda}. In particular we are considering $M=75000 \; m$. In \fref{conditional wave function-figure1} the numerical solution of the squared modulus of $\Psi(x_1,x_2,t)$ is plotted at four different times. At the initial time $t = 0$,  \fref{conditional wave function-figure1}~(a),  the entire wave function is at the left of the barrier.  At a later time $t_0$ the wave function has split up into reflected and transmitted parts due to the barrier, see \fref{conditional wave function-figure1}~(b).  Then, because the electron has not yet arrived at the transmitted charge detector, the wave function has the following form: 

\begin{equation}
\Psi(x_1,x_2,t_0) = \left[\psi_{T}(x_1,t_0)+\psi_{R}(x_1,t_0)\right]\phi(x_2,t_0).
\end{equation}

After that, Figs. \ref{conditional wave function-figure1} (c) and (d), the
interaction of the detector with the transmitted part of the wave function appears. For time $t>t_0$ the transmitted part of the wave
function is shifted up in the $x_2$ direction while the reflected part does not move. The interaction with the apparatus thus produces 
two channels in the configuration space, one corresponding to the electron being transmitted and the other corresponding to the electron being
reflected, getting an entangled superposition among the electron and the apparatus.

\begin{figure}
\centering
\includegraphics[width=0.97\columnwidth]{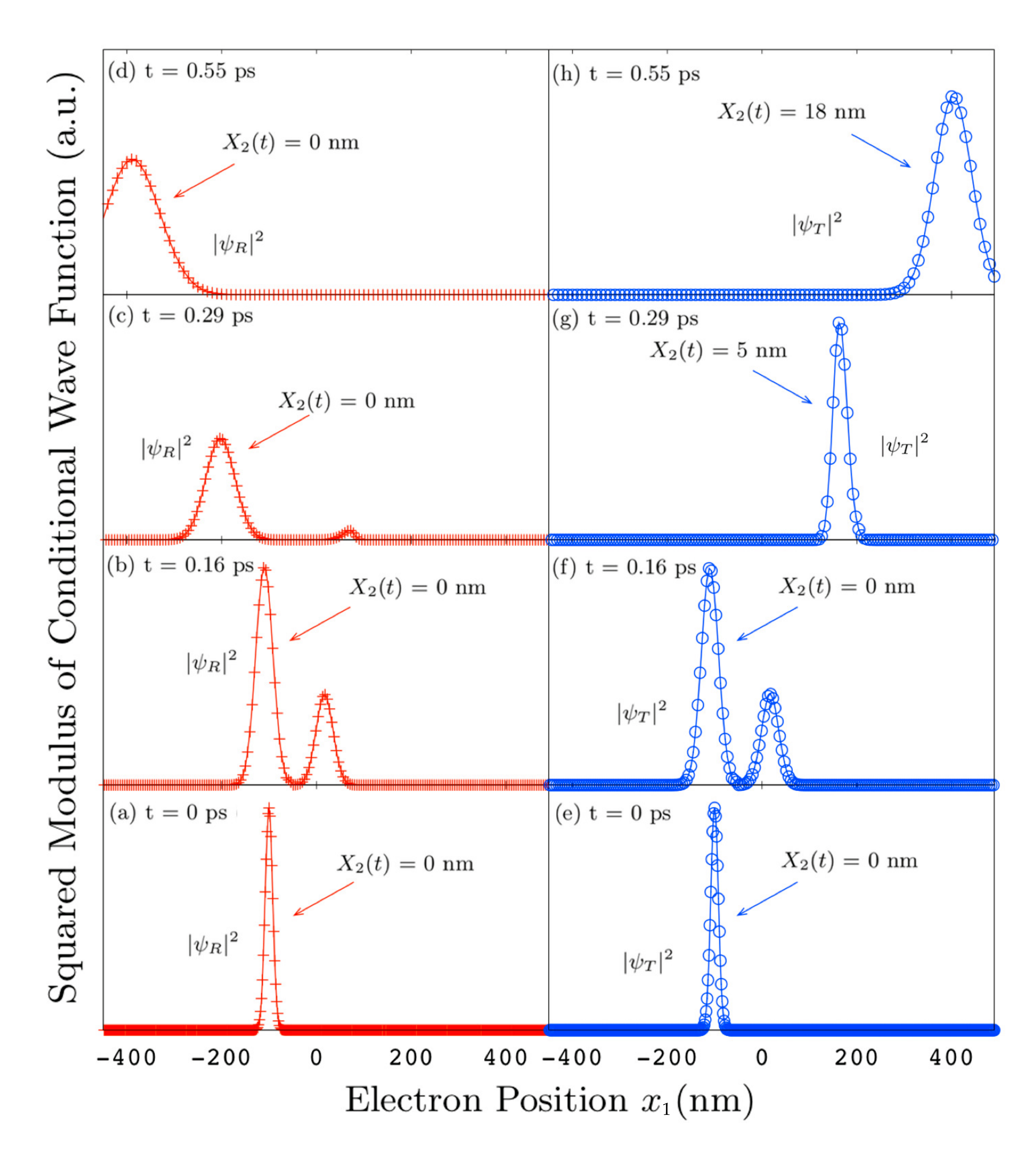}
\caption{ The $+$ line in (a), (b), (c) and (d) is the time evolution of the squared modulus of the conditional wave function associated to the trajectory $\alpha = 1$ in \fref{conditional wave function-figure1}, i.e. $\psi_{R} = |\Psi(x_1,X_2^{\alpha=1}(t),t)|$. The $\odot$ line in (e), (f), (g) and (h) is the squared modulus of the conditional wave function associated to the trajectory $\alpha = 3$ in \fref{conditional wave function-figure1}. i.e. $\psi_{T} = |\Psi(x_1,X_2^{\alpha=3}(t),t)|$. The actual detector position $X_2(t)$ is plotted at each time in order to compare these results with those in \fref{conditional wave function-figure1} (Color figure online)}
\label{conditional wave function-figure2}
\end{figure}

In \fref{conditional wave function-figure1} we also plot the actual positions of the system and detector $\{X_1(t),X_2(t)\}$ for four different possible initial positions $\{X_1(0), X_2(0)\}$, corresponding (say) to four distinct runs of the experiment (labelled by $\alpha=1,...,4$). Of the four possible evolutions shown, three show the electron being transmitting ($\alpha=2,3,4$) and one being reflecting ($\alpha=1$). While the pointer position $X_2(t)$ does not move for the reflected particle, its evolution for the transmitted ones clearly shows a movement. In conclusion, looking at the \emph{detector} position we can perfectly certify if the particle has been reflected ($X_1(t)< -50\;nm$ and $X_2(t)=0\;nm $) or transmitted ($X_1(t)> -50\;nm$ and $X_2(t) \approx 15\;nm $). We hope the reader will realize how trivially we have been
able to explain the measurement, using only a \emph{channelized} (unitary) time-evolution of 2D wave function plus two Bohmian trajectories, one for the system and another for the measuring apparatus. 

Once we have solved the complete problem of the measurement  in (2D) configuration space, we can describe the same measurement in (1D) physical space with the help of the conditional wave function. The key point illustrated here is that the collapse of the one-particle wave function for the electron, which collapse is of course postulated through the second law seen in \sref{sec2} in ordinary quantum theory, instead arises naturally and automatically in Bohmian mechanics.  It is simply a consequence of slicing the unitary-evolving (2D) wave function $\Psi$ along the (moving) line $x_2 = X_2(t)$, resulting $\psi_1(x_1,t)=\Psi(x_1,X_2(t),t)$. In \fref{conditional wave function-figure1} we have plotted two solid horizontal lines corresponding to a slice of the wave function at two different values of $X_2(t)$. In \fref{conditional wave function-figure2} we report the evolution of these (time-dependent) slices of the many-particle wave function, the \emph{conditional wave function} for the electron, for the trajectories $\alpha = 1$ and $\alpha = 3$ from \fref{conditional wave function-figure1}. We clearly see that if the particle is reflected, as it is the case for $\alpha = 1$, the position of the pointer does not change with time and, after the interaction with the detector has been performed, the electron's \emph{conditional wave function} includes only a reflected part. See Figs. \ref{conditional wave function-figure2} (c) and (d). On the other hand, when the particle is transmitted (e.g., $\alpha =3$), it is the reflected part of the \emph{conditional wave function} which collapses away, leaving only the transmitted packet.  See Figs. \ref{conditional wave  function-figure2} (g) and (h). Note in particular that the 1D evolution of $\psi_1(x_1,t)$ (the electron's \emph{conditional wave function}) is not unitary, even though the 2D evolution of $\Psi$ is. 

While a wave function formulation of quantum mechanics provides only statistical information about the experimental results, with the help of the Bohmian trajectories, we have been able to recover the individual result of each experiment. In fact, for each experiment the pointer of the detector is either moving (corresponding to a transmitted electron) or not (reflected electron), while an ensemble of repeated experiments (where the initial positions of the particles, both the electron $X_1(0)$ and the detector $X_2(0)$, are selected according to the squared modulus of the wave function at the initial time $|\Psi(x_1,x_2,0)|^2$) reproduce the same statistical results.

Thus with the previous numerical example we have reproduced the collapse-behaviour of the wave function of a transmitted (or reflected) electron. This allows us to conclude that the same results of standard formalism (explained in \sref{sec2}) are obtained within Bohmian mechanics (see \cite{OriolsBook,Goldstein1} for a formal derivation of the empirical equivalence of the two theories). Apart from irrelevant technicalities (related to how we define the measuring apparatus) the results in \fref{figure2} and \fref{conditional wave function-figure2} are conceptually identical. We emphasize that, the collapse in Bohmian theory is naturally derived. Such a natural derivation of the collapse behavior demystifies the measurement process (and the quantum noise). We underline that we achieve the non-unitary evolution of the wave function of a measured system simply slicing the enlarged wave function (which includes the apparatus) in the configuration space.

Let us return to the questions posed at the end of \sref{sec2} about the measurement of the total (conduction or displacement) current. Is this measurement process \emph{continuous} or \emph{instantaneous}? Does it provide a \emph{strong} or a \emph{weak} perturbation of the wave function? The Bohmian theory does not provide simple answers to these questions, but it clearly indicates the path.  We need to include (somehow) the measuring apparatus in the Hamiltonian. Here, the electrostatic interaction between the electrons in the system and those in the contacts, cables, etc. See a preliminary work in this direction in Ref. \cite{IWCE-Paris}. 

A powerful simulator which uses Bohmian mechanics to compute DC and quantum noise is the BITLLES simulator \cite{xavier2,BITLLES}. This simulator allows us to work with a lot of flexibility, being able of introducing any sort of potential, including Coulomb correlations and exchange interaction. The next Section is dedicated to expose the fundamental ideas of the simulator and an example of the calculation of noise with it.

\section{Practical application}
\label{sec4}

We have previously exposed the main features of Bohmian mechanics explaining in a quite trivial way the partition noise in a tunneling barrier. However, as it occurs for all theories, there is a huge step between its general formulation and its practical application. In fact, it happens many times that the practical problem we want to solve is unsolvable both analytically and numerically, and some kind of approximations are required. The paradigmatic example of the need of approximations in quantum theories is the well-known \emph{many-body problem} that reminds us that the celebrated Schr\"odinger equation in \eref{TDSE} (or any alternative formulation) can only be solved exactly for very few (one, two, three,..) degrees of freedom \cite{Manybody,Dirac}. 

In principle we have seen in \sref{sec3} that to reproduce the collapse of the wave function in Bohmian mechanics we have to include a suitable interaction with an external apparatus. Then we can write down the Schr\"odinger equation for our complete system including all the electrons in the active region of the device plus all the particles composing the detector. But solving numerically this problem is obviously impossible. Again, the \emph{many-body problem} appears. Then we should look for suitable approximations able to reduce the complexity of our problem. Let us emphasize that the (technical) approximations that we will show, do not alter the general framework we have previously presented.

\subsection{An approximation for the interaction between the electron and the measuring apparatus}
\label{sec4.1}

The first kind of approximation regards the inclusion of the apparatus in our simulations. It seems that its inclusion is unavoidable in order to provide the collapse of the wave function. And this is true, but in the particular case of quantum noise in electrical devices, the fact of \emph{playing} with (Bohmian) trajectories will greatly simplify the problem. In \aref{app-bohm-3}, we have reported how any experimental value is calculated in Bohmian mechanics. The important thing is that any expectation value of a given operator is simply calculated as a function of the actual particles positions over an ensemble of repeated experiments (see \eref{mean-value} in \aref{app-bohm-3}). Thus what really matters in the computation of a property of the quantum system are only the trajectories of the Bohmian particles (not the wave functions). Therefore, if the trajectories without measuring apparatus are enough accurate (this means if the error on these trajectories due to neglecting the apparatus is reasonably small compared to the exact solution) we can get accurate results with a minimal computational effort. In the case of the transmitted charge detector of \sref{sec3}, it has been demonstrated \cite{albareda} that the error due to the exclusion of the apparatus from the simulations is almost negligible for the computation of the trajectories. In this way we can decrease enormously the computational burden, removing all the degrees of freedom related to the apparatus from our computations. 

We can provide a more didactic discussion on why the previous technical approximation for the measuring apparatus works quite well when using Bohmian trajectories. In \sref{sec2}, we conclude that the reason why the wave function evolution in \fref{figure1} was wrong is due to the wrong possibility that an electron that is transmitted at time $t_1$ is later reflected at time $t_2$. This unphysical result simply disappears when using Bohmian trajectories: the dynamic of a transmitted electron at time $t_1$ will be determined \emph{locally} by the guidance law \eref{eq-guidance} that only takes into account the transmitted part of the wave function. We can, for all practical purposes, completely ignore the reflected part of the wave function. Therefore, at time $t_2$, this electron will remain as a transmitted electron with full certainty.\\    

Finally, let us emphasize that, in principle, the measuring apparatus has also a role in the classical simulation of electronic devices. Such interaction with the apparatus is included at a classical level, at best, by a proper boundary conditions for the scalar potential of the Hamiltonian (i.e. the Poisson equation) ensuring overall charge neutrality. Obviously, this kind of approximation can also be included here.

\subsection{An approximation for the Coulomb and exchange interaction among electrons}
\label{sec4.2}

Once we have \emph{practically} eliminated the apparatus from our computations, a second kind of approximation regards the interactions among the electrons of our device. The active region of the electronic device can contain hundreds of electrons. Also in this case, as we mentioned, the many-particle Schr\"{o}dinger equation can be solved only for very few degrees of freedom. A standard way to proceed consists then on reducing the complexity of the problem by \emph{tracing out} certain degrees of freedom. This process ends up with what is called the \emph{reduced density matrix}. When the reduced density matrix is used, its equation of motion is no longer described by the Schr\"odinger equation but in general by a non-unitary operator. The reduced density matrix is no longer a pure state, but a mixture of states and their evolution is in general irreversible \cite{DiVentra}. Now we discuss how Bohmian mechanics allows us to proceed in a very different way. As it will be seen below, the concept of \emph{conditional wave function} \cite{Goldstein1} provides an original tool to deal with many-body open quantum systems \cite{xavier1,Nino1}. \\

As said, once again the key instrument is the conditional wave function. In order to use the conditional wave function to reduce the degrees of freedom of a system we must know how it evolves in time. It can be demonstrated \cite{xavier1} that the single-particle conditional wave function of particle $1$, $\psi_{1}(x_1, t)$, for a system of $N$ interacting particles, obeys the following wave equation:
\begin{eqnarray}\label{Conditional_eq}
i\hbar\frac{\partial \psi_1(x_1,t)}{\partial t}=\Big\{-\frac{\hbar^2}{2m} \nabla^2_1+U_{1}(x_1,\bar{X}_{N-1}(t),t) \nonumber \\ +G_{1}(x_1,\!\bar{X}_{N\!-\!1}(t),t)\!+\!i J_{1}(x_1,\!\bar{X}_{N-1}(t),t) \Big\} \psi_{1}(x_1,t).
\end{eqnarray}
The explicit expression of the potentials $G_{1}(x_1,\!\bar{X}_{\!N\!-\!1\!}(t),t)$ and $J_{1}(x_1,\bar{X}_{N-1}(t),t)$ that appears in \eref{Conditional_eq} can be found in reference \cite{xavier1}. However, their numerical values are in principle unknown and need some educated guesses.
On the other hand, the total electrostatic potential energy among the $N$ electrons that appears in \eref{TDSE}, has been divided into two parts:
\begin{eqnarray}\label{Potential_energy}
U(x_1,\bar{X}_{N-1}(t),t)&=&U_{1}(x_1,\bar{X}_{N-1}(t),t)+ \nonumber \\ 
&+&U_{N-1}(\bar{X}_{N-1}(t),t).
\end{eqnarray}
The term $U_{1}(x_1,\bar{X}_{N-1}(t),t)$ can be any type of many-particle potential defined in the position-representation, in particular it can include short-range and long-range Coulomb interactions. The other term $U_{N-1}(\bar{X}_{N-1}(t),t)$ in \eref{Potential_energy} without dependence on $X_1$, is contained in the coupling potential $G_1$ in \eref{Conditional_eq}. The same procedure can be done for the rest of the $N-1$ particles, for example for particle $2$ we fix the positions of particle $1,3,...,N$ obtaining the analogous of \eref{Conditional_eq} for $\psi_2(x_2,t)$. From a practical point of view, all quantum trajectories $\bar{X}_N(t)$ have to be computed simultaneously. In order to gather all the above concepts, let us discuss a practical computation with conditional wave functions by detailing a sequential procedure:
\begin{enumerate}
\item At the initial time $t=0$, we fix the initial position of all $i$-particles, $X_i(0)$, according to the initial probability distribution ($|\Psi(\bar{x}_N,0)|^2$), and their associated single-particle wave function $\psi_{i}(x_i,0)$.
\item From all particle positions, we compute the exact value of the potential $U_{i}(x_i,\bar{X}_{N-1}(0),0)$ for each particle. An approximation for the terms $G_{i}$ and $J_{i}$ is required at this point. We use the simplest one \cite{albareda}.
\item We then solve each single-particle Schr\"{o}dinger-type equation, \eref{Conditional_eq}, from $t=0$ till $t=dt$.
\item From the knowledge of the single-particle wave function $\psi_{i}(x_i, dt)$, we can compute the new velocities $v_{i}(dt)$ for each $\!i$-particle (see \eref{eq-guid-cwf} in \aref{app-bohm-2}).
\item With the previous velocity, we compute the new position of each $i$-particle as $X_{i}(dt)=X_{i}(0)+v_{i}(dt)dt$.
\item Finally, with the set of new positions and wave functions, we repeat the whole procedure (steps 1 till 5) for another infinitesimal time $dt$ till the total simulation time is finished.
\end{enumerate}
The advantage of the above algorithm using \eref{Conditional_eq} instead of the many-particle Schr\"odinger equation (\eref{TDSE}) is that, in order to find approximate trajectories, $X_i(t)$, we do not need to evaluate the wave function and potential energies in the whole configuration space, but only over a smaller number of configuration points, $\{\!x_i,\! \bar{X}_{\!N\!-\!1\!}(t)\!\}$, associated with those trajectories defining the highest probabilities according to $|\psi(\bar{x}_N,t)|^2$ . \\

The exchange interaction is naturally included in \eref{Conditional_eq} through the terms $G_{i}$ and $J_i$. Due to the Pauli exclusion principle, the modulus of the wave function tends to zero, $|\psi(x_i, \bar{X}_{\!N\!-\!1\!}(t),t)|\to 0$, in any neighborhood of $x_{i}$ such that $|x_{i}-\bar{X}_{N-1}(t)|\to 0$. Thus, both terms, $G_{i}(x_i, \bar{X}_{\!N\!-\!1\!}(t),t)$ and $J_{i}(x_i, \bar{X}_{\!N\!-\!1\!}(t),t)$, have asymptotes at ${x_{i}}\to {\bar{X}_{N-1}}(t)$ that \textit{repel} the $i-$ particle from other electrons. However, in order to exactly compute the terms $G_i$ and $J_i$ we must know the total wave function, which is in principle unknown. There are however a few ways to introduce the symmetry of the wave function without dealing directly with these two coupling terms \cite{xavier1,aalarcon09pps,alarconUnpublished}. Clearly, the complexity of the algorithm increases as we go beyond the single-particle quantum transport scenario mentioned in \sref{sec2} and \sref{sec3}.

\subsection{Practical example}
\label{practical}

An electron device is an open system, where many parameters can only be estimated from the knowledge of their statistical (typical) distribution. Apart from the uncertainty in the initial position in the quantum trajectories (the $\alpha$ distribution explained in \aref{app-bohm}), we also take into account the uncertainty on the properties of the injected electrons (initial energies, momentums, etc) which we refer to the parameter $h$. The random process $I^{\gamma}(t)$ mentioned in \sref{sec1.1} is now written as $I^{\alpha,h}(t)$. At finite temperature, the thermal noise introduces fluctuations on the energies of the electrons entering inside the device. As discussed in the introduction of \sref{sec2}, the study of the noise in electrical devices due to the partition noise of the barrier plus the thermal noise of the injection are traditionally known as quantum shot noise \cite{Landauer1,Beenakker1,Beenakker2,Buttiker3,Buttiker2,Landauer3}. This is the noise studied in this \sref{sec4}. In many systems, one obtains the well known Schottky's result \cite{Schottky} or Poissonian shot noise, $S_{II\;shot}(0)=2q\left\langle I\right\rangle$, for the noise power spectral density defined in \eref{noise5} at zero frequency, i.e. $w=0$.\\

We select a particular (large) set of wave packets with values $\alpha$ and $h$ for selecting their initial conditions. We refer to such selection as $\alpha_1$ and $h_1$. We evolve the wave packets and trajectories as explained in previous paragraphs. Within the approximation mentioned in \sref{sec4.1} and \sref{sec4.2}, the total current value can be calculated as the sum of the particle or conduction current plus the displacement current:

\begin{eqnarray}
I^{\alpha,h}(t) & = & I_c^{\alpha,h}(t)+I_d^{\alpha,h}(t)= \nonumber \\
& = & \int_{S} \sum_{i=1}^N q_i v_i(X_i^{\alpha,h}(t))\delta(x_S-X_{i}^{\alpha,h}(t)) \cdot ds + \nonumber  \\
&+& \int_{S} \sum_{i=1}^N\epsilon(x_S)\dfrac{dE(x_S;X_i^{\alpha,h}(t),t)}{dt} \cdot ds,
\label{current}
\end{eqnarray}

where $S$ is the surface where we want to calculate the current, $x_S$ are the points of the chosen surface, $\epsilon(x_S)$ is the dielectric constant in the same surface and $E(x_S;X_i^{\alpha,h}(t),t)$ is the electric field in the surface $S$ which depends on the actual position of all the electrons. 

Once we know $I^{\alpha_1,h_1}(t)$ for a large interval of time, the algorithm to compute the current fluctuations is quite simple following \eref{noise3} and \eref{noise5}. This discussion can be familiar for those people who works in semi-classical approaches. In fact, the Bohmian procedure explained here for quantum transport is very similar to that of, for instance, the Monte-Carlo simulations of the Boltzmann equation. But instead of being the electric-field the one who \emph{guides} the electrons, it is the wave function, through the guiding velocity field in \eref{eq-guidance}. 

\begin{figure}
\centering
\includegraphics[width=0.97\columnwidth]{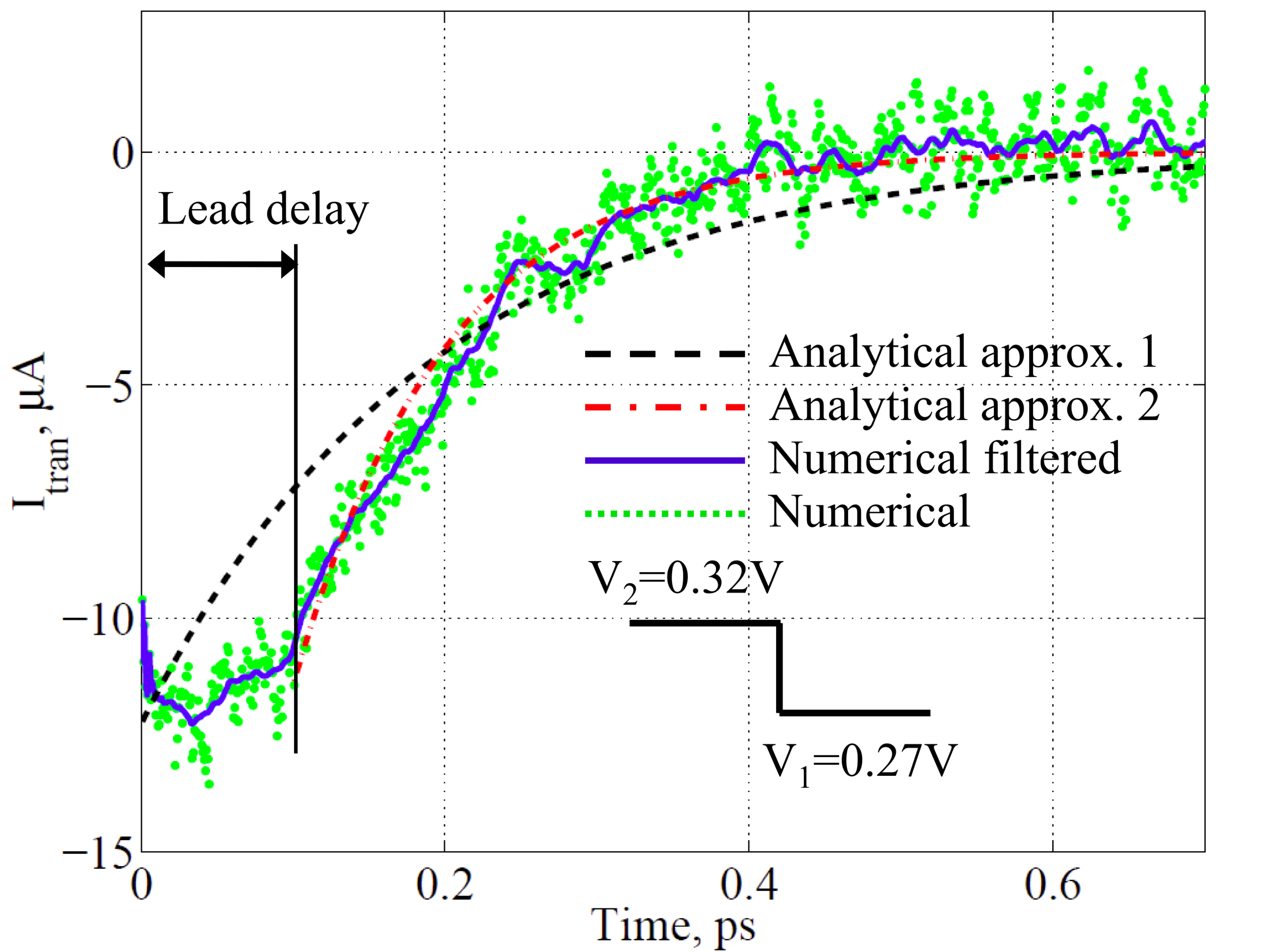}
\caption{Transient current $I_{tran}(t)$ computed analytically and numerically (Color figure online)}
\label{F6_alarcon:Figure6_9a}
\end{figure}

As a practical example of the computation of the fluctuations, we show here the current response to a step input voltage in the Negative Differential Conductance region of a RTD. The input signal is the step voltage $V(t)=V_{1}u(t)+V_{2}\left[  1-u(t)\right]$ where $u(t)$ is the Heaviside (step) function. The voltages $V_{1}$ and $V_{2}$ are constant. Then the current response can be expressed as $I(t)=I_{tran}(t)+I_{1}u(t)+I_{2}\left[  1-u(t)\right]$ where $I_{1}$ and $I_{2}$ are the stationary currents corresponding to $V_{1}$ and $V_{2}$ respectively, and $I_{tran}$ is the \textit{intrinsic} transient current. 

\begin{figure}
\centering
\includegraphics[width=0.97\columnwidth]{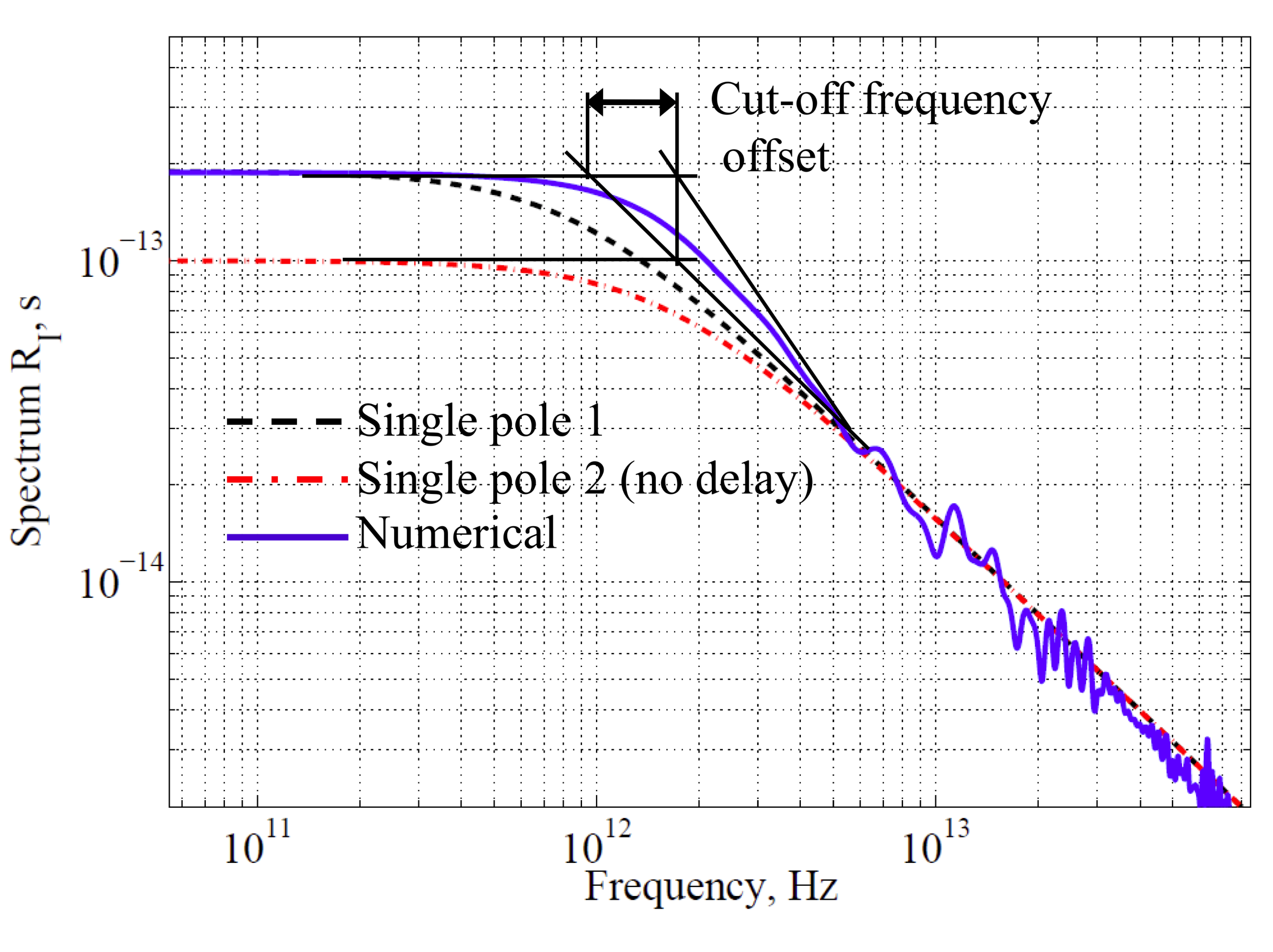}
\caption{Fourier transform of $I_{tran}(t)$ of \fref{F6_alarcon:Figure6_9a}. A logarithmic scale is used to resolve the cut-off frequency offset (Color figure online)}
\label{F6_alarcon:Figure6_9b}
\end{figure}

The results are reported in \fref{F6_alarcon:Figure6_9a} where $I_{tran}(t)$ manifests a delay with respect to the step input voltage, due to the dynamical adjustment of the electric field in the conductors. After this delay, the current response becomes a RLC-like response (dot-dashed line RLC response $2$) i.e. purely exponential. Performing the Fourier transform of $I_{tran}(t)$ in \fref{F6_alarcon:Figure6_9b} and comparing it with the single pole spectra (Fourier transform of the RLC-like responses, dashed and dashed dotted lines), we are able to estimate the cut-off frequency and the frequency offset due to the delay \cite{xavier2}.

\begin{figure}
\includegraphics[width=0.97\columnwidth]{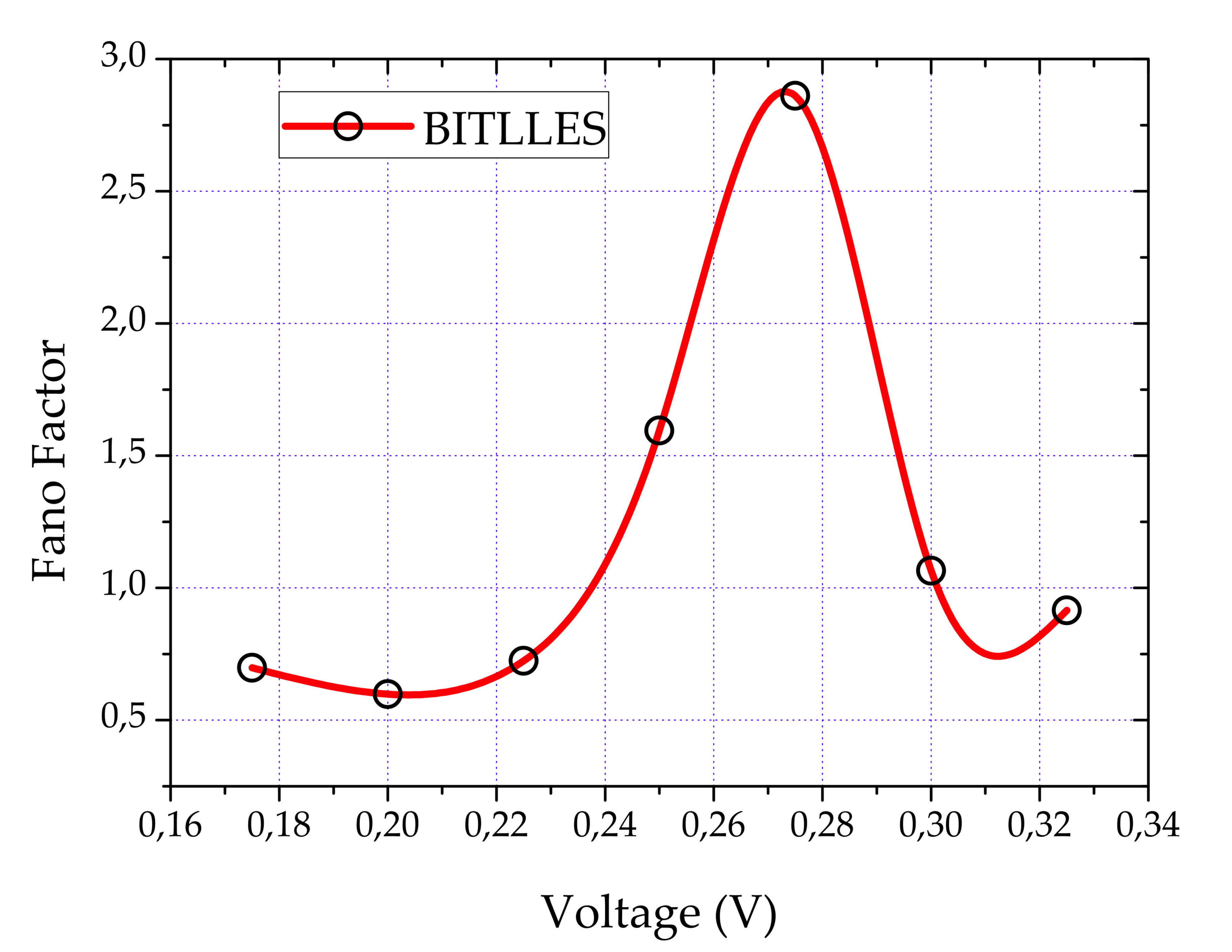}
\caption{Fano Factor $F$ defined as $F=S(0)/(2q\left\langle I\right\rangle)$, evaluated using the current fluctuations directly available from the BITLLES simulator (Color figure online)}
\label{F6_alarcon:Figure6_10}
\end{figure}

\begin{figure}
\centering
\includegraphics[width=0.97\columnwidth]{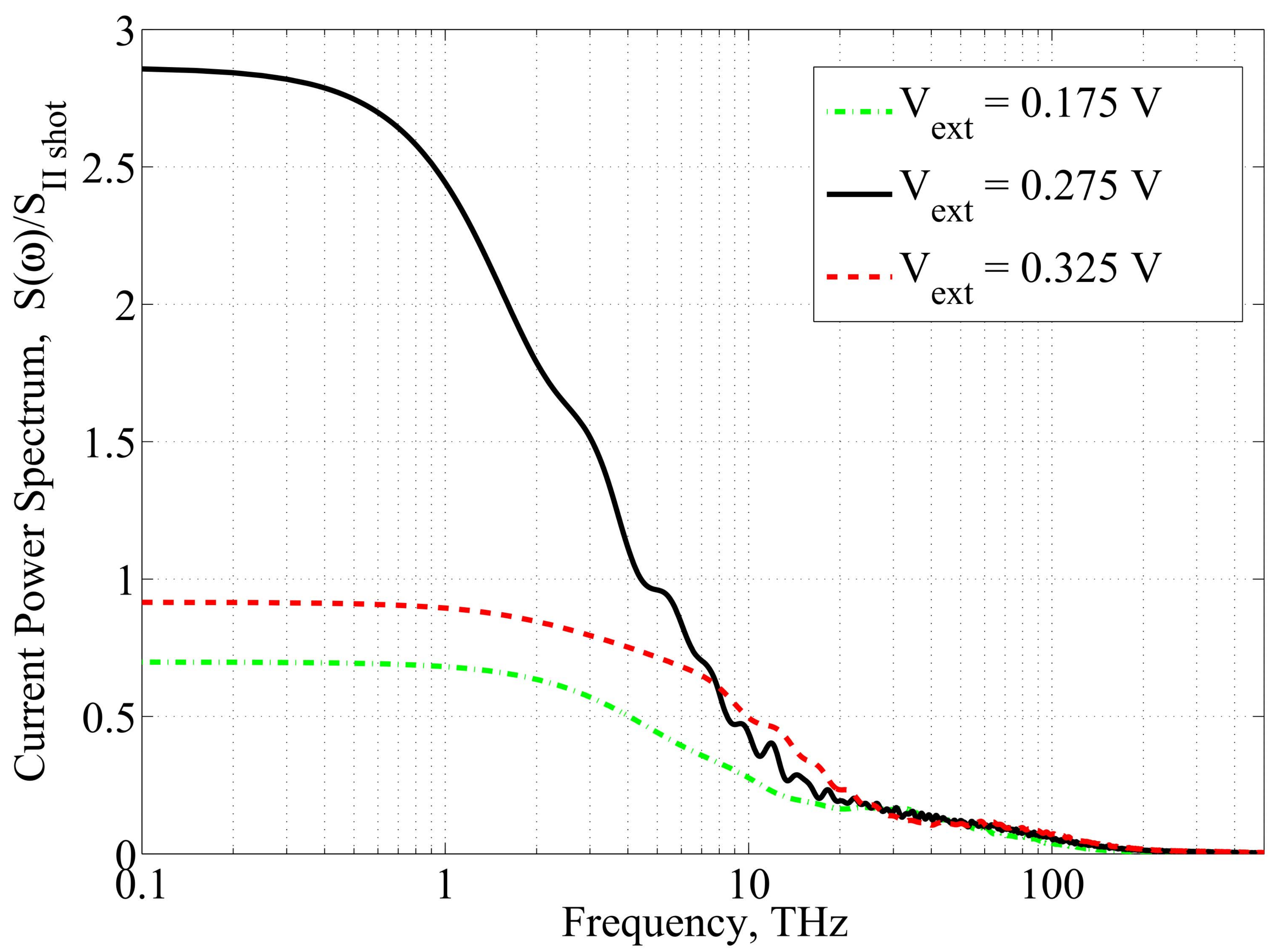}
\caption{Current noise power spectrum referred to Poissonian shot noise at different biases (Color figure online)}
\label{F6_alarcon:Figure6_12}
\end{figure}

In order to understand how the many-body Coulomb interaction affects the noise in RTDs, we also investigate the correlation between an electron trapped in the resonant state during a dwell time $\tau_{d}$ and those remaining in the left reservoir.
This correlation occurs essentially because the trapped electron perturbs the potential energy felt by the electrons in the reservoir. In the limit of non-interacting electrons, the Fano factor will be essentially proportional to the partition noise, however, if the dynamical Coulomb correlations are included in the simulations (see \fref{F6_alarcon:Figure6_10}) this result is no longer reached, it shows super-poissonian values. Finally, we are also interested in the high frequency spectrum $S(w)$ given by \eref{noise5} revealing information about the internal energy scales of the RTD that is not available from DC transport (see \fref{F6_alarcon:Figure6_12}).

\section{Conclusions}
\label{sec5}

In the present paper we discuss the understanding and the computation of quantum noise in electrical devices from a Bohmian perspective. Computations of quantum noise are quite complex because noise is generally quantified in terms of temporal correlations. Such correlations must include the time-evolution of a quantum system during and after a measurement. Usually, many other quantum computations do only require a final measurement, so that their time-evolution from the initial until the final time is uniquely determined by the unitary (Schr\"odinger like) evolution. As discussed in \fref{figure2}, this unitary evolution is not enough to compute time correlations which require mixing unitary and non-unitary (the so-called collapse of the wave function) time evolutions.  

There are several (empirically equivalent) quantum theories. Each quantum theory has its own formalism that is able to connect the experimental values with some abstract elements such as wave functions, operators, trajectories, etc. that are able to satisfactorily reproduce (or predict) experimental results. We discuss how the Copenhagen (also known as standard or orthodox) interpretation and Bohmian mechanics give explanation to the partition noise. For a flux of electrons impinging upon a tunneling barrier, we analyze how a measurement process affects partition noise in a quantum device. For simplicity, to focus on the importance of the measurement process, we consider spinless electrons without Coulomb and exchange interaction.  In \sref{sec2} we explain how standard quantum theory provides an answer for the measurement problem by means of the introduction of the notion of operators. We see that this notion is not always satisfactory even for practical purposes, because the definition of which is the \emph{right} operator is not obvious. Then in \sref{sec3} we discuss an alternative way to deal with the collapse without introducing the idea of operators. In fact within Bohmian mechanics, a theory of wave \emph{and} particles, the collapse is derived trivially by means of the introduction of the \emph{conditional wave function} (the wave function of a subsystem), a tool exclusively belonging to Bohm's theory. Obviously, each theory gives a different formalism to compute quantum noise and different interpretation of its origin. In any case, at the end of the day, the same empirical predictions are achieved by using the orthodox quantum theory or Bohmian mechanics.  

In \sref{sec4}, because the Bohmian formulation uses trajectories to compute experimental results, we see that a very reasonable approximation to include collapse can be achieved with a very small computational effort. Finally, details of the simulator named BITLLES based on Bohmian mechanics and numerical results for low and high frequency noise of the current in a resonant tunneling diode are presented. We emphasize that the presented formalism and the procedure for computing the properties of a system (in our case current, noise, etc.) have many similarities with the one used in semi-classical simulations (for example Monte-Carlo of the Boltzmann equation \cite{tomas}). In any case, Bohmian formalism is not at all a semi-classical approach but a complete quantum theory that can be applied to study any non-relativistic quantum phenomena, quantum noise and collapse among them. 

Finally, we wish to discuss which is the ultimate origin of the quantum noise according to orthodox and Bohmian interpretations. Before entering into details, let us recall that the definition of noise given in \eref{noise2} in \sref{sec1} is just the difference between the experimental value $I(t)$ and what we define as the signal. Therefore, even a sinusoidal current $I(t)$ provides a value of $\triangle I^2$ different from zero. What we want to discuss hereafter are not all the possible sources of fluctuations in $I(t)$, but only if there is any new type of randomness in quantum devices that is not present in classical ones and what is its origin.  

Given this last specification, we can provide an answer to the question: \emph{What is the ultimate origin of quantum noise according to the orthodox interpretation?} As we see in \sref{sec2}, the transmission or reflection of a single electron impinging upon a tunneling barrier becomes unpredictable. This is an example of a new source of randomness present only in quantum devices. According to the orthodox theory, this randomness appears because of the collapse of the wave function due to measurements. Without the collapse (that is put in by hand as an additional postulate in the orthodox theory), the wave function follows a deterministic law dictated by the Schr\"odinger equation. The partition noise in the tunneling barrier discussed in \sref{sec2} is due to the action of the operator which implements the random collapse of the wave function (selecting the final wave function stochastically among the set of available eigenstates). 

Alternatively, we can also answer the question: \emph{What is the ultimate origin of quantum noise according to Bohmian mechanics?} The randomness in the values of the current in $I(t)$ provided by the BITLLES simulator comes from the $h$ and  $\alpha$ distributions mentioned in \sref{practical}. The $h$ distribution is due to the uncertainty of the initial energies, momentums, entering times, etc. of the electrons. This source of ``extrinsic'' randomness can be minimized imagining technological improvements of the setup (for example, well-controlled single electron sources). On the contrary, the $\alpha$ distribution of the conditional wave function (explained in \aref{app-bohm}) is an unavoidable source of randomness. This randomness of the $\alpha$ distribution cannot be minimized by any technological improvement \cite{wiseman,acin}. Therefore, whether the particle is transmitted or reflected becomes unpredictable in Bohmian theory too. Thus, though Bohmian mechanics is deterministic,  an \emph{appearance} of randomness emerges in the subsystems \cite{Goldstein1,Zanghi_chaos}. It is important to notice that the measurement of the system does not introduce any additional randomness.  The ultimate origin of the unpredictability is the fact that the uncertainty principle does not allow us to known the (well-defined in the Bohmian theory) initial position of the particles in each experiment.

In summary, according to the orthodox interpretation, the partition noise has its origin in the stochasticity of the orthodox measurement process. On the contrary, Bohmian mechanics says that the origin of noise is the uncertainty of the initial position of the trajectory in each realization of the experiment. Although both theories give the same predictions, in the authors' opinion, the latter has a more natural, common and understandable explanation of the origin of quantum noise. While following deterministic laws, the transmission or reflection of a Bohmian electron is unpredictable in a given experiment. A classical dice is a simpler example of a system following deterministic laws that becomes unpredictable.  Collapse in Bohmian mechanics is so naturally derived that the quantum measurement problem, in general, and quantum noise, in particular, are somehow demystified. We underline that Bohmian mechanics achieves the non-unitary evolution of the wave function of a measured system simply slicing the enlarged wave function in the configuration space (without introducing any measurement-associated randomness).

We accept that preferences between the explanation of the origin of quantum noise in terms of the orthodox or Bohmian interpretations are subjective. Therefore, in this paper we have also developed objective arguments about the computational advantages of the Bohmian formalism. The facts that the measuring apparatus, what we call the ammeter, is directly included into the Hamiltonian of the Schr\"odinger equation and that the current values are computed from trajectories (not from the wave functions) allow us to study system plus apparatus scenarios (or look for reasonable approximations). This ability is very relevant, for example, in the computation of high frequency currents where it is difficult to find the \emph{right} operator. For all these reasons, we conclude that quantum noise is easily understood and computed from a Bohmian perspective in many practical scenarios.   

\begin{acknowledgements}
We want to acknowledge T. Norsen, N. Zangh\`i, G. Albareda and F.L. Traversa for insightful discussions.This work has been partially supported by the \lq\lq{}Ministerio de Ciencia e Innovaci\'{o}n\rq\rq{} through the Spanish Project TEC2012-31330 and by the Grant agreement no: 604391 of the Flagship initiative  \lq\lq{}Graphene-Based Revolutions in ICT and Beyond\rq\rq{}. Z. Z acknowledges financial support from the China Scholarship Council (CSC). D.M. is supported in part by INFN and acknowledges the support of COST action (MP1006) through STSM.
\end{acknowledgements}

\begin{appendices}
\section{The quantum DC current in ergodic systems}
\label{app-noise}
The DC current measured in a laboratory $\langle I \rangle$ can be computed by time-averaging the measured value of the total current $I(t)$  from a \emph{unique device} during a large (ideally infinite) period of time $T$ as mentioned in \eref{noise1}. If we can justify the ergodicity of electronic devices, we can alternatively compute $\langle I \rangle$  from an ensemble-average of all possible values of the current $I_i$ measured, at one particular time  $t$, over an \emph{ensemble of (identical) devices} as seen in \eref{noise7}. For DC quantum transport computations, Eq. (\ref{noise7}) is greatly preferred because it deals directly with the probabilistic interpretation of the wave function. It is important to realize that while \eref{noise1} implies measuring the quantum current many times, \eref{noise7} involves only one measurement.  We do not need to worry about the evolution of the wave function after the measurement when using \eref{noise7}. Let us discuss this point in more detail. We can define the eigenstates $| \psi_i \rangle$ of a particular operator $I$, as those vectors that satisfy the equation $I |\psi_i \rangle = I_i | \psi_i \rangle $. The eigenvalue $I_i$ is one of the $M$ possible measured values in \eref{noise7}.\footnote{For simplicity we assume that there is no degeneracy. Our qualitative discussion does not change if degeneracy is considered.} Since the entire set of eigenstates form a basis of the Hilbert space, the wave function can be decomposed as $| \psi (t) \rangle = \sum_{i=1}^M c_i(t)  |\psi_i \rangle  $, with $c_i(t) = \langle \psi_i | \psi(t) \rangle$. Then, we can rediscover \eref{noise7} as follow:

\begin{eqnarray}
\langle I \rangle &=& \langle \psi(t)| I | \psi(t) \rangle = \nonumber \\
&=& \sum_{j=1}^M c^{*}_j(t) \langle \psi_j | \sum_{i=1}^{M} I_ic_i(t) |\psi_i\rangle = \sum_{i=1}^{M} I_i P(I_i),
\end{eqnarray} 
where we have used the orthonormal property of the eigenstates $\langle \psi_j | \psi_i \rangle = \delta_{ij}$ and the definition of the (Born) probability $P(I_i) \!=\! |c_i(t)|^2$. We emphasize that $\langle\! \psi(t)| I | \psi(t)\! \rangle$ does not require the explicit knowledge of the eigenstates. Only the free evolution of the state $| \psi(t) \rangle$ and the \emph{measuring} operator $I$ are needed.  

At this point, it is mandatory to provide some discussion about the use of the ergodic theorem. Strictly speaking, no ergodic theorem exists for an out of equilibrium system \cite{Price}. Indeed, an out of equilibrium system is represented by a distribution function, or probability function, that is different from that in equilibrium and arises from a balance between the driving forces and the dissipative forces. The applied bias used to measure the DC current of any device implies that the device is quite likely in a far from equilibrium state. Therefore, the ergodic connection between Eq. (\ref{noise1}) and Eq. (\ref{noise7}) has to be considered as only a very reasonable approximation for DC transport, but not as an exact result \cite{Price}.

\section{Bohmian mechanics}
\label{app-bohm}

Bohmian mechanics is a version of quantum theory whose basic elements are waves and point-like particles. The many-particle wave function evolves according to the Schr\"odinger equation (\ref{TDSE}) while particles have definite position at any time with a law given by \eref{eq-guidance}, therefore being a fully deterministic theory. The configuration of the particles, say at time $t=0$, is chosen randomly according to $|\Psi|^2$ at the initial time, known as \emph{quantum equilibrium hypothesis} \cite{Goldstein1}. Thanks to the continuity equation

\begin{eqnarray}
\label{eq-continuity}
\frac{\partial \rho}{\partial t} = - \nabla \left( \rho v \right),
\end{eqnarray}

where $\rho = |\Psi|^2$ and $v$ the Bohmian vector field. An important consequence of the quantum equilibrium hypothesis and equivariance is the empirical equivalence between Bohmian mechanics and orthodox quantum theory for any kind of non-relativistic quantum experiments.

\subsection{The conditional wave function}
\label{app-bohm-2}

Consider a quantum system of $N$ particles and a partition of it in such a way that its spatial coordinates can be split as $\bar{x}_N = \{x_1, \bar{x}_{N-1}\}$. Where we denote with $x_1$ the position in $\mathbb{R}^3$ space of the electron $1$, while with $\bar{x}_{N-1}$ the positions of the rest of the electrons in a $\mathbb{R}^{3 (N-1)}$ space. The actual particle trajectories are accordingly denoted by $\bar{X}_N(t) = \{X_1(t), \bar{X}_{N-1}(t)\}$. \emph{How can one assign a wave function to the electron $1$?} In general this is not possible if the two subsystems are entangled, i.e. the total wave function cannot be written as a product $\Psi(\bar{x}_N) = \psi_1(x_1)\psi_{N-1}(\bar{x}_{N-1})$. However, we can modify our question and ask what is the wave function of the electron $1$ that provides the exact velocity $v_1$ given a particular configuration $\bar{X}_{N-1}(t)$ for the rest of the particles. The answer given by Bohmian mechanics is the so called \emph{conditional wave function} \cite{Goldstein1,Nino1}:
\begin{equation}
\psi_{1}(x_1,t) = \Psi(x_1,\bar{X}_{N-1}(t),t),
\label{Conditional}
\end{equation}
which constitutes a slice of the whole multi-dimensional wave function. The wave function constructed in such a way gives exactly the same Bohmian velocity 

\begin{eqnarray}
v_1(t) \!=\! \frac{\hbar}{m_1} \text{Im} \frac{\nabla_1 \Psi}{\Psi}\Big|_{\bar{x}_N\!=\!\bar{X}_N(t)}\! \!\equiv\! \frac{\hbar}{m_1} \text{Im} \frac{\nabla_1 \psi_1}{\psi_1}\Big|_{x_1 \!=\! X_1(t)}.
\label{eq-guid-cwf}
\end{eqnarray}

\subsection{Computation of mean value of an operator}
\label{app-bohm-3}

If needed, Bohmian mechanics can make use of operators, but only as a mathematical trick. Without any physical or fundamental role in the operator. We briefly explain how it is possible to calculate the mean value of a general hermitian operator with Bohmian trajectories. The \emph{quantum equilibrium hypothesis} at the initial time $t=0$ can be expressed in terms of the trajectories as follows

\begin{eqnarray}
\label{QEH}
|\Psi(\bar{x}_N,0)|^2 = \lim_{M_{\alpha} \to \infty} \frac{1}{M_{\alpha}} \sum_{\alpha=1}^{M_\alpha} \prod_{i=1}^{N} \delta(x_i-X_i^{\alpha}(0)),
\end{eqnarray}  

where the superindex $\alpha$ takes into account the uncertainty in the initial position of the particles. It can be easily demonstrated \cite{OriolsBook} that the evolution of the above infinite set of quantum trajectories $\alpha=1,2,...,M_{\alpha}$ reproduce at any time $t$ the probability distribution, $|\Psi(\bar{x}_N,t)|^2$. \\
For computing the mean value of an operator $A$ it can be demonstrated \cite{OriolsBook} that 

\begin{eqnarray}
\langle A \rangle_{\Psi} = \lim_{M_{\alpha} \to \infty} \frac{1}{M_{\alpha}} \sum_{\alpha=1}^{M_\alpha} A_B(\bar{X}_N^{\alpha}(t)),
\label{mean-value}
\end{eqnarray}

where $A_B(\bar{x}_N)$ is the ``local" mean value of $A$.

\end{appendices}

\end{document}